\documentstyle[prl,aps,floats,epsf]{revtex}\def\narrowtext{}\tighten\twocolumn
\input epsf.sty
\begin{document}
\draft

\title{
Zeeman Perturbed $^{63}$Cu Nuclear Quadrupole Resonance Study of
the Vortex State of
YBa$_2$Cu$_3$O$_{7-\delta}$
}
\author{
Y. Itoh,$^{1}$ C. Michioka,$^{1}$ K. Yoshimura,$^{1}$ and Y. Ueda,$^{2}$
}

\address{
$^1$Department of Chemistry, Graduate School of Science,\\
Kyoto University, Kyoto 606-8502, Japan \\
$^2$Institute for Solid State Physics, 
University of Tokyo, 5-1-15 Kashiwanoha,\\
Kashiwa, Chiba 277-8581, Japan\\
}

\date{\today}%
\maketitle %

%\address{\begin{minipage}[t]{6.0in} %
\begin{abstract} 
We report a $^{63}$Cu nuclear quadrupole resonance (NQR) study of the vortex state 
for an aligned polycrystalline sample 
of a slightly overdoped high-$T_c$ superconductor
YBa$_2$Cu$_3$O$_{7-\delta}$ ($T_{c}\sim$92 K) 
at a low magnetic field of 96 mT along the $c$ axis,
near a lower critical field $H_{c1}$.  
We observed 
the frequency distribution of the nuclear spin-lattice relaxation time $^{63}T_1$ 
in the Zeeman-perturbed $^{63}$Cu NQR spectrum 
below $T_c$. 
The characteristic behavior of 1/$^{63}T_1$, 
taking the minimum values with respect to temperature and frequency,   
indicates the significant role of antiferromagnetic spin fluctuations 
in the Doppler-shifted quasiparticle energy spectrum 
inside and outside vortex cores. 
%\typeout{polish abstract} %
\end{abstract}
%\end{minipage}} %

%\maketitle %
\narrowtext 

Strong correlation effects on the electronic state inside and outside vortex cores
in the mixed state of type-II superconductors have attracted great interests,
because an interplay between magnetic correaltion and unconventional
superconductivity could be observed at quasi-normal cores~\cite{SCZhang}. 
Scanning tunneling spectroscopy~\cite{STM} and muon spin rotation technique~\cite{mSR} have been applied
to study the local quasiparticle density of states and the local field distribution
around the vortex cores, respectively. 
The site-selective NMR techniques have also been applied to study 
the magnetism and the charged states of the vortex cores 
of high-$T_c$ cuprate superconductors~\cite{Curro,Mitro,Kaku}.
For high magnetic fields, however, the intervortex transfer of quasiparticles and 
the disorder of the vortex lattice prevent us
from observing the original state of an isolated vortex core.  
For low magnetic fields, the well separated, large vortex cores~\cite{uSR}
may allow us to observe the original magnetism of a single vortex core.  
To our knowledge, no NMR study on the local electron spin dynamics inside and outside vortex cores
near a lower critical field $H_{c1}$ 
has been reported so far for the high-$T_c$ superconductors. 

YBa$_2$Cu$_3$O$_{7-\delta}$ (Y123) is a widely known high-$T_c$ superconductor. 
The optimal $T_c$ is about 93 K for an oxygen content of $\delta$=$0.06-0.08$. 
The zero-field plane-site Cu nuclear quadrupole resonance (NQR) spectrum is one of the sharpest 
among the reported spectra of other cuprate superconductors.  
The Cu nuclei are coupled with nearly uniaxial electric field gradients~\cite{Shimizu}.
For Y123,
the hysteresis curves of magnetization below $T_c$ 
indicate various types of the vortex pinning effect at high magnetic fields~\cite{Erba}. 
At a low magnetic field near $H_{c1}$, 
a large diamagnetic response is observed in any case. 
The vortex lattice of Y123 at a low magnetic field is in a quasi-long range order 
and stable in a broad temperature range~\cite{Koba}.  
One can obtain direct information of the antiferromagnetic spin correlation 
on the CuO$_2$ planes through the Cu NMR/NQR. 
In the Cu NMR experiment for Y123, 
the low magnetic field yields only a negligibly small Cu Knight shift 
but is expected to cause a large diamagnetic shift. 
In passing, a low field NMR experiment was recently performed 
for a superconducting Sr$_2$RuO$_4$
to estimate a large Ru Knight shift
but not the diamagnetic shift~\cite{KIshida}.  
  
In this Letter, we report a Zeeman-split plane-site $^{63}$Cu NQR study of the vortex state 
of a slightly overdoped high-$T_{c}$ superconductor Y123 ($T_{c}\sim$92 K) 
at a low magnetic field of 96 mT along the $c$ axis. 
The sample is a $^{63}$Cu-isotope enrich powder, 
already used in Ref.~\cite{Itoh}, 
mixed with Stycast 1266 epoxy and magnetically aligned along the $c$ axis. 
An intervortex distance at 96 mT is $\sim$1800 $\AA$, 
slightly longer than the penetration depth 
$\lambda_{ab}$($T\rightarrow$0 K)=1100-1300 $ \AA$~\cite{uSR}. 
For $\delta\sim$0.05, $H_{c1}$ is about 110 mT at $T\rightarrow$0 K~\cite{Hc1}.
We observed the effect of superconducting diamagnetic field distribution on the split NQR spectrum,
the characteristic temperature and frequency dependence of 
the $^{63}$Cu nuclear spin-lattice relaxation time $^{63}T_{1}$ below $T_c$. 
It turned out 
that the antiferromagnetic spin fluctuations 
develop at the vortex cores on a dilute vortex lattice. 
    
A phase-coherent-type pulsed spectrometer was utilized to perform 
the NMR/NQR experiments.   
All the NMR measurements were done while cooling in a magnetic field of 96 mT. 
The pure NQR spectra and the Zeeman-split frequency spectra 
were measured with quadrature detection, 
where integrations of the nuclear spin-echoes were recorded
as functions of frequency $\nu$ 
while $\nu$ was changed point by point.
The power spectra of the spin-echo signals will be shown below, 
because the spectra include weak signals over a broad frequency region. 
Throughout the present measurements,  
the frequency window excited by rf pulses was $\nu_1\sim$83 kHz, 
which was estimated 
from the first exciting $\pi$/2-pulse $t_{w}=3 \mu$s and
the relation of $\pi/2=2\pi\nu_{1}t_{w}$.  
The nuclear spin-lattice relaxation curves
$p(t)\equiv 1-M(t)/M(\infty)$ (recovery curves) 
of the nuclear spin-echo amplitude $M(t)$
were measured by an inversion recovery technique,
as functions of a time $t$ after an inversion pulse.  
 
\begin{figure}
\epsfxsize=2.4in
\epsfbox{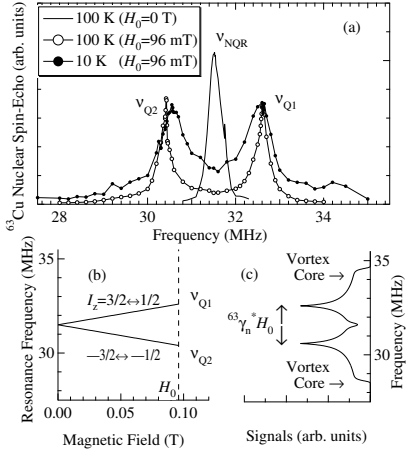}
\vspace{0.0cm}
\caption{
(a) Zeeman-split $^{63}$Cu (a spin $I$=3/2) NQR spectra at $T$=10 and 100 K. 
(b) Zeeman-split $^{63}$Cu NQR lines (solid lines with $\nu_{\mathrm{Q1}}$ and $\nu_{\mathrm{Q2}}$) as functions of
an external magnetic field along the $c$-axis.   
The dashed line is $H_0$=96 mT. 
(c) A model illustration of the diamagnetic field distribution of 
the Zeeman-split NQR spectrum of a spin $I$=3/2. 
The scale of the spectrum is not realistic. 
The actual diamagnetic shift is superimposed on the original broad NQR lines.  
}
\label{CuNMR}
\end{figure}

 Figure 1(a) shows a zero-field $^{63}$Cu NQR power spectrum at 100 K ($>T_c$) and 
the Zeeman-split NQR spectra at 100 K and 10 K ($<T_c$).  
The magnetic field of $H_{0}$=96 mT was applied along the $c$ axis, 
the maximum principle axis of electric field gradients. 
The applied magnetic field splits the $^{63}$Cu NQR spectrum with a peak frequency $\nu_{\mathrm{NQR}}$
into two lines with lower and higher frequency peaks 
($\nu_{Q1}$ and $\nu_{Q2}$)~\cite{DasHahn,Takigawa1}.
Using definition of $^{63}\gamma_n^{\ast}$$\equiv$(1+$K_c$)$^{63}\gamma_n$ 
($^{63}\gamma_n$=11.285 MHz/T and the $c$-axis Knight shift $K_c$),
the transition lines of $\nu_{Q1}$ ($I_z$=3/2$\leftrightarrow$1/2) and 
$\nu_{Q2}$ ($I_z$=-3/2$\leftrightarrow$-1/2) are expressed by 
 
\begin{eqnarray}
\left\{
\begin{array}{l}
\nu_{\mathrm{Q1}}=\nu_{\mathrm{NQR}}+^{63}\gamma_{n}^{\ast}H_{0},\\
\nu_{\mathrm{Q2}}=\nu_{\mathrm{NQR}}-^{63}\gamma_{n}^{\ast}H_{0}.
\end{array}
\right. 
\label{e.ZPNQR}
\end{eqnarray}
One should note that $K_c$ includes not only the Knight shift but also the superconducting diamagnetic shift. 
Figure 1(b) illustrates a diagram 
of the resonance frequency against the magnetic field.  
From the observed $\nu_{Q1}$ and $\nu_{Q2}$,  
one can estimate the quadrupole frequency $\nu_{\mathrm{NQR}}$($H$) at a magnetic field $H$ 
and the local magnetic field $\delta h$[$\equiv$$(1+K_{\mathrm{c}})H_{0}$] 
by

\begin{eqnarray}
\left\{
\begin{array}{l}
\nu_{\mathrm{NQR}}(H)=(\nu_{Q1}+\nu_{Q2})/2, \\
\delta h=(\nu_{Q1}-\nu_{Q2})/2{}^{63}\gamma_{n}.
\end{array}
\right. 
\label{e.eqQKs}
\end{eqnarray} 
The origin of the linewidth at 100 K is primarily electric,
i.e. the distribution of the electric field gradients.  
The split $^{63}$Cu NQR lines broaden at 10 K. 
The local field due to the Knight shift of 1.27 $\%$~\cite{Takigawa2} 
is negligible at the magnetic field of $H_{0}$=96 mT.  
Therefore, the line broadening at 10 K must be due to the superconducting diamagnetic shift
$^{\mathrm{sc}}K_{c, \mathrm{dia}}$ or 
due to the distribution of the electric field gradient $\nu_{\mathrm{NQR}}(H)$
in a mixed state.  

\begin{figure}
\epsfxsize=2.1in
\epsfbox{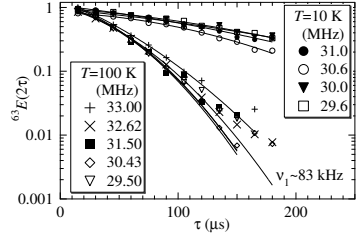}
\vspace{0.0cm}
\caption{
Frequency distribution of the transverse relaxation curve of
$^{63}$Cu nuclear spin-echo $^{63}E(2\tau)$ at $T$=10 and 100 K. 
The solid curves are fits by Gaussian-times-exponential functions
$E(2\tau)$=$E(0)$exp[-(2$\tau/T_{2L}$)-0.5(2$\tau/T_{2G}$)$^2$]
for eye guides.
} 
\label{CuDecay}
\end{figure}

In order to determine whether the broadening at 10 K is magnetic or electric, 
we measured the frequency dependence of 
$^{63}$Cu nuclear spin-echo decay curves $^{63}E(2\tau)$ 
($\tau$ is a delay time between $\pi$/2- and $\pi$-pulses) at 10 K and 100 K.  
As shown in Fig. 2, 
the transverse relaxation is slower at 10 K than at 100 K
and no appreciable frequency dependence of $^{63}E(2\tau)$ is observed.  
The slower transverse relaxation in the superconducting state 
indicates the occurrence of a detuning effect on the nuclear spin-spin coupling~\cite{Silbernagel}. 
The local field distribution due to the vortices shifts the resonance frequency of the neighboring nuclei
out of the excited frequency region of $\nu_{1}\sim$83 kHz. 
A part of the $\it like$-spin (resonant) nuclei at 100 K becomes the $\it unlike$-spin (nonresonant) ones at 10 K. 
The in-plane Cu nuclear spin-spin coupling is expressed 
by $a_{ij}I_{zi}I_{zj}$ ($i$ and $j$ are the nuclear sites),
where $a_{ij}$ is  
enhanced via an antiferromagnetic spin susceptibility~\cite{PS}. 
The slow nuclear spin-spin relaxation in Fig. 2 
is not consistent with the zero field Cu NQR results~\cite{Itoh},
and then it should not be intrinsic 
in the local antiferromagnetic correlation.  
The detuning effect on the transverse relaxation confirms
that the diamagnetic field distribution results in the additional broadening at 10 K.  
 
First, we estimated 31.54 $\leq\nu_{\mathrm{NQR}}(\mathrm{96 mT})\leq$31.58 MHz
at 10 K. 
We have already measured the $T$-dependence of $\nu_{\mathrm{NQR}}(0)$, 
e.g. $\nu_{\mathrm{NQR}}$=31.54 MHz at 10 K.
We estimated the upper limit of the field-induced quadrupole frequency 

\begin{equation} 
0\leq[\nu_{\mathrm{NQR}}(96 \mathrm{mT})-\nu_{{NQR}}(0)]\leq 40 \mathrm{kHz}. 
\label{e.DeqQ}
\end{equation}
This is slightly smaller than the reported values for charged vortices at 9.4 T~\cite{VCKuma}. 
The low magnetic field of 96 mT does not enhance the vortex charge for Y123.  

Second, we estimated $\delta h$=90$\pm$1 mT at 10 K. 
The value of $\delta h$ is smaller than the applied field $H_0$=96 mT,
so that the peaks of the split lines are affected by negative shift of -6 mT. 
One may suppose the Redfield pattern in the field distribution 
superimposed on the split NQR lines.  
The peak lines are regarded as the saddle points in the diamagnetic field distribution
on a vortex lattice~\cite{TIM,Riseman,Brandt,Red}. 
The highest and the lowest frequency edges are regarded as the vortex core positions. 
In this assignment, the site-selective NMR at a low field is possible for the present Y123 after Ref.~\cite{Curro}. 
The superconducting parameters of the penetration depth $\lambda$ and the coherence length $\xi$, 
however, could not be estimated from Fig. 1(a),  
because it is difficult to model the inhomogeneous NQR lines
and then the actual distribution of diamagnetic field.  
The line profile of the split NQR spectrum at 10 K in Fig. 1(a)
just imitates the Redfield pattern 
and results predominantly from the inhomogeneous distribution of electric field gradients.

\begin{figure}
\epsfxsize=3.2in
\epsfbox{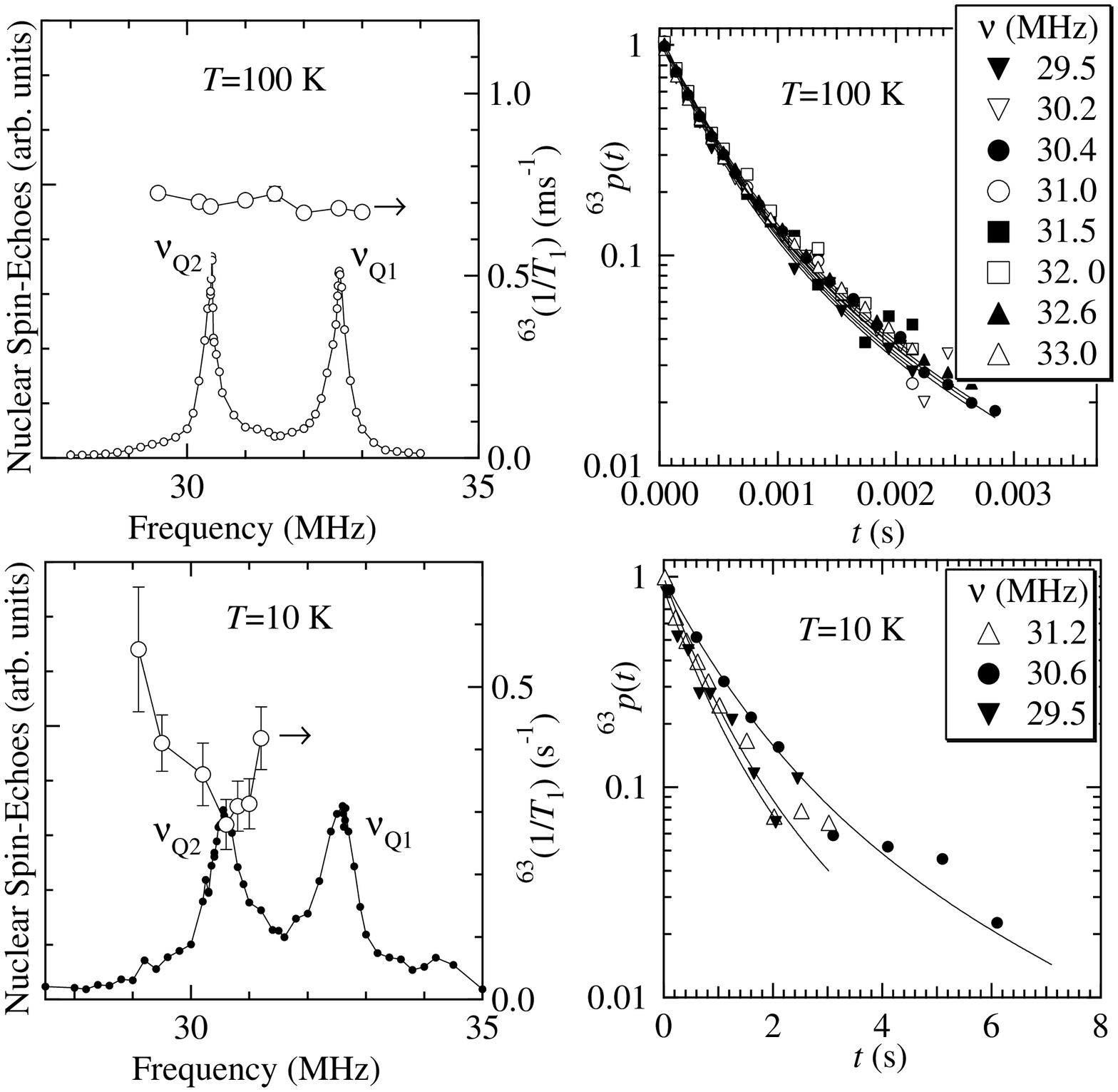}
\vspace{0.0cm}
\caption{
Frequency distribution of
$^{63}$Cu nuclear spin-lattice recovery curves (right panels)
in the Zeeman-perturbed Cu NQR spectra (left panels) 
at $T$=100 K (upper panels) and $T$=10 K (lower panels). 
The solid curves in the right panels are best fits by eq. (4). 
The frequency dependence of 1/$^{63}T_{1}$ are shown in the left panels. 
}
\label{CuT1}
\end{figure}

Figure 3 shows
the frequency distribution of
$^{63}$Cu nuclear spin-lattice recovery curves $^{63}p(t)$ (right panels)
in the Zeeman-perturbed Cu NQR spectra (left panels) 
at $T$=100 K (upper panels) and $T$=10 K (lower panels). 

The solid curves in the right panels of Fig. 3
are least-squares fits to the recovery curves $^{63}p(t)$ by 

\begin{equation} 
p(t)=p(0)[0.1\mathrm{e}^{-t/{T_1}}+0.5\mathrm{e}^{-3t/{T_1}}+0.4\mathrm{e}^{-6t/{T_1}}], 
\label{e.reco}
\end{equation}
where $p(0)$ and $T_1$ are the fitting parameters. 
Equation (\ref{e.reco}) is derived from a rate equation with 
a magnetic transition between the nuclear spin states $I_z$=$\pm$3/2$\leftrightarrow$$\pm$1/2~\cite{Takigawa1}. 
No large deviation from eq. (\ref{e.reco}) was observed below $T_c$. 

The frequency dependences of the $^{63}$Cu nuclear spin-lattice relaxation rate 1/$^{63}T_{1}$ 
are shown in the left panels of Fig. 3.  
At 100 K ($> T_c$), 1/$^{63}T_{1}$ shows no appreciable frequency dependence,
whereas at 10 K ($< T_c$), it shows a strong dependence. 
This is an evidence for the site-selective measurement of 1/$^{63}T_{1}$ 
in the Zeeman-split NQR spectrum 
and for the less effect of nuclear spin diffuison.

The frequency dependence of 1/$^{63}T_{1}$ 
is similar to those of the plane-site $^{17}$O(2, 3) and the apical $^{17}$O(4) 
at high magnetic fields of $H_{0}$=9-37 T~\cite{Curro,Mitro}. 
The plane-site Cu directly probes the in-plane antiferromagnetic correlation,
but the plane-site oxygen does not. 
The frequency dependence of 1/$^{17}T_1$ is understood primarily 
by change in quasiparticle excitations around the vortex cores 
due to Doppler shift~\cite{TIM,Morr}. 
The supercurrent around a vortex core induces Doppler shift 
in the local quasiparticle energy spectrum~\cite{deGennes}. 
Therefore, the frequency dependence of 1/$^{63}T_{1}$, 
similar to that of the high field 1/$^{17}T_{1}$,
indicates the importance of the Doppler-shifted quasiparticle energy spectrum and
the absence of the static antiferromagnetic vortex cores at 96 mT. 
Coexistence of vortex solid with liquid is reported at higher magnetic fields~\cite{Reyes}. 
The coexistence could be expected just below $T_c$ at a low field
but was not clear in this study. 

\begin{figure}
\epsfxsize=3.3in
\epsfbox{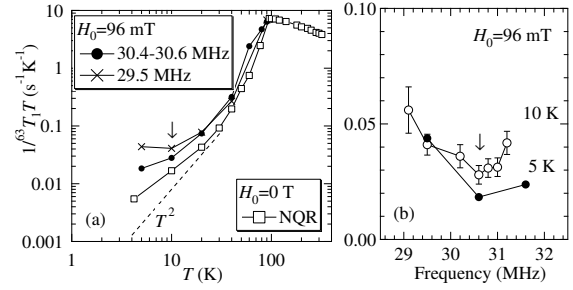}
\vspace{0.0cm}
\caption{
(a) Temperature dependence of 
the $^{63}$Cu nuclear spin-lattice relaxation rate divided by temperature 1/$^{63}T_{1}T$ 
across the split NQR spectrum
and at the zero field NQR frequency.
The arrow indicates a minimum of 1/$^{63}T_{1}T$ at 10 K
for the 29.5 MHz nuclei close to the vortex cores.
(b) Frequency dependence of 1/$^{63}T_{1}T$ at $T$=5 and 10 K. 
The arrow indicates a minimum of 1/$^{63}T_{1}T$
for nuclei close to the saddle points in a vortex lattice. 
}
\label{CuT1}
\end{figure}

Figure 4(a) shows 1/$^{63}T_{1}T$ 
as a function of temperature
across the split NQR spectrum
and at the zero field NQR frequency. 
The signals at 30.4-30.6 MHz and at 29.5 MHz in the split NQR spectrum come from the nuclei
close to the saddle points and close to the vortex cores, respectively.
The zero field 1/$^{63}T_{1}T$ is consistent with the results in Ref.~\cite{Imai}.   
The dashed line is a function of 1/$T_{1}T\propto T^{2}$ due to a $d$-wave superconducting gap with line nodes. 
The temperature dependence of 1/$^{63}T_{1}T$ at each selected site 
can be understood primarily by the spatial dependence of 
the local density of states with the Doppler-shifted quasiparticle energy spectrum
~\cite{TIM}.  
The minimum of 1/$^{63}T_{1}T$ at 10 K for nuclei at 29.5 MHz, however,
is not understood simply by the local density of states. 

Figure 4(b) shows 1/$^{63}T_{1}T$ 
as a function of frequency at $T$=5 and 10 K. 
The increase in 1/$^{63}T_{1}T$ from the saddle points to the vortex cores 
can be understood by the Doppler-shifted quasiparticle energy spectrum.
But, the minimum behavior around the saddle points is not understood
only by the Doppler shift effect.
 
The superconducting order is locally destroyed by the field-induced supercurrent,
and then the competing fluctuations develop in the vortex cores. 
The supercurrent-induced scattering process via the antiferromagnetic spin fluctuations
in the Doppler-shifted quasiparticle energy spectrum plays a significant role
around the vortex cores~\cite{Morr}.   
The local antiferromagnetic spin fluctuations causes competition  
between thermal quasiparticle scattering and the creation/annihilation of two quasiparticles 
in a spin-fermion model~\cite{Morr}. 
The minimum behavior of 1/$^{63}T_{1}T$ with respect to temperature and frequency
is an evidence for such a competition mechanism inside and outside the vortex cores
with the antiferromagnetic spin fluctuations. 

To conclude, the effect of antiferromagnetic spin fluctuations  
on the low-lying excitation inside and outside vortex cores
was observed in the $^{63}$Cu nuclear spin-lattice relaxation measurements
for Y123 at 96 mT near $H_{c1}$. 

\acknowledgments
We thank Dr. M. Kato for his experimental supports.   
This work was supported in part by Grants-in Aid for Scientific Research
 from Japan Society for the Promotion of Science (16076210).
% references

\end{document}